\documentclass[aps,pra,reprint,superscriptaddress]{revtex4-1}
\usepackage{graphicx}
\usepackage{amsmath}
\usepackage{bm}
\usepackage[alsoload=synchem]{siunitx}
\usepackage{braket}
\usepackage{epstopdf}
\usepackage{url}
\usepackage[utf8]{inputenc}
\DeclareUnicodeCharacter{2212}{-}

\begin{document}
	
	\title{Dual-colour magic-wavelength trap for suppression of light shifts in atoms}
	
	\author{A. P. Hilton}
	\author{C. Perrella}
	\author{A. N. Luiten}
	\author{P. S. Light}
	\email[]{Philip.Light@Adelaide.edu.au}
	\affiliation{Institute for Photonics and Advanced Sensing (IPAS) \& School of Physical Sciences, The University of Adelaide, Adelaide, South Australia 5005, Australia}
	
	\date{\today}

\begin{abstract}
	We present an optical approach to compensating for spatially varying ac-Stark shifts that appear on atomic ensembles subject to strong optical control or trapping fields.
	The introduction of an additional weak light field produces an intentional perturbation between atomic states that is tuned to suppress the influence of the strong field.
	The compensation field suppresses sensitivity in one of the transition frequencies of the trapped atoms to both the atomic distribution and motion.
	We demonstrate this technique in a cold rubidium ensemble and show a reduction in inhomogeneous broadening in the trap.
	This two-colour approach emulates the magic trapping approach that is used in modern atomic lattice clocks but provides greater flexibility in choice of atomic species, probe transition, and trap wavelength.
\end{abstract}

\pacs{}

\maketitle

\section{Introduction}
	Cold atomic ensembles are broadly applied in the highest precision sensing and measurement applications.
	These samples have nearly ideal properties including strong interaction with light for measurement and preparation, finely tuned internal interactions, as well as high environmental isolation and long internal state lifetimes.
	As such, atom ensembles are being applied to quantum logic operations \cite{Lo2011, Feizpour2015, Liu2016}, quantum memories \cite{Sparkes2013, Cho2016, Blatt2016, Ma2017, Hsiao2018}, optical atomic clocks \cite{Takamoto2005, Targat2013, Bloom2014}, atomic interferometers \cite{Hardman2016}, and tests of fundamental physics \cite{Fixler2007, Rosenband2008, Bouchendira2011}.
	To make the most of the advantageous properties of cold atomic ensembles it is frequently necessary to trap and position the ensemble \cite{Grimm2000}.
	This optical manipulation necessarily alters the energies of the internal atomic states to produce the attractive potential, which in turn results in a distribution of light-induced shifts of the transition frequencies as atoms sample the intensity distribution of the trapping field.
	Frequently one is forced into making a trade-off between the benefits of a tight and deep trap and the undesirable shifting and broadening of the atomic states that are used in the experiment.
	
	The solution to this problem, developed in state-of-the-art atomic lattice clocks, has been termed the magic wavelength trap \cite{McKeever2003, Ye2008, Katori2009}.
	In this case, a careful choice of trap wavelength can eliminate unwanted shifts in a particular transition.
	This effect is tailored to be independent of trap intensity by creating equal light-shifts in the two states connected by the transition.
	This technique has become universally adopted and a number of magic wavelengths have been calculated and/or demonstrated for a variety of atomic species \cite{Katori2003, Arora2007, Barber2008, Hinkley2013, Wang2016}.
	
	Despite its revolutionary impact in creating the ultimate high-performance atomic clocks, the magic technique is not universally applicable.
	Many commonly used species do not possess a suitable energy level structure, while for others the necessary wavelengths and powers for the trapping field may not be viable.
	
	In such cases, more complex solutions have been developed where auxiliary magnetic fields or particular polarisation states are used \cite{Flambaum2008,Singh2016}.
	In other circumstances, it has been suggested that bichromatic traps might offer a solution \cite{Griffin2006,Arora2010,metbulut2015a}.
	Indeed, some calculated possibilities for Rb and Cs have been published \cite{Arora2010,Wang2014,metbulut2015a}.
	Nonetheless, to our knowledge no-one has yet demonstrated a bichromatic magic trap for an optical transition.
	
	We present a simple theoretical proscription of the bichromatic magic wavelength technique and demonstrate it in a four-level system in a cold rubidium ensemble.
	We provide a numerical model using the density matrix formalism and show excellent agreement between the experimental results and the simulation.
	This opens the benefits of the magic wavelength technique for cancelling spatially dependent light shifts to a much broader array of atomic species and transitions.

\section{Methods}
	\begin{figure}
		\centering
		\includegraphics{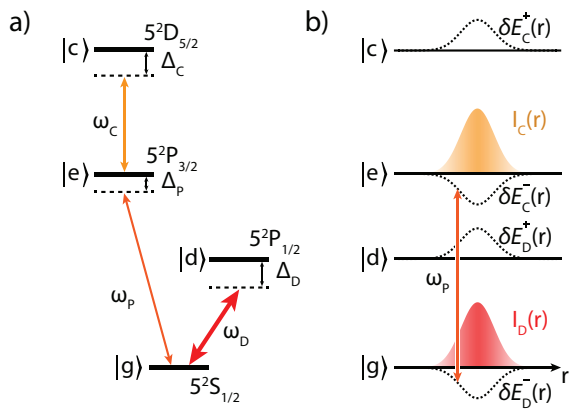}
		\caption{
			\label{fig:EnergyLevels}
			Diagrams of 
			a) the relevant energy levels and applied optical fields with frequency detunings shown explicitly, and 
			b) the spatially varying light shifts induced by the dipole and compensation fields on the targeted states in the desired magic regime.
		}
	\end{figure}
	We implement this technique on dipole-trapped Rb atoms.
	The Rb is trapped  using the $5\mathrm{S}_{1/2} \rightarrow 5\mathrm{P}_{1/2}$ $\mathrm{D}1$ line, labelled $\ket{\mathrm{g}} \rightarrow \ket{\mathrm{d}}$, see in Figure \ref{fig:EnergyLevels}.a.
	For a light field detuned far from an optical transition such as this, the energy levels are simply shifted symmetrically in lower $(\delta E^-)$ and upper $(\delta E^+)$ states.
	The magnitude of this shift in energy is given by the product of the reduced Planck’s constant and the generalised Rabi frequency, $\hbar \Omega$, and, for a far-detuned dipole field, the shift in $\ket{\mathrm{g}}$ and $\ket{\mathrm{d}}$ is found by a series expansion around $\Omega/\Delta \rightarrow 0$:
	\begin{equation}
		\label{eqn:EnergyShift}
		\delta E^\pm = \mp \hbar \Omega \approx \mp \frac{1}{2 \epsilon_0 c \hbar} \left| \bra{\mathrm{g}} \bm{\mu} \ket{\mathrm{g}} \right|^2 \frac{I_\mathrm{D}\left(\mathbf{r}\right)}{\Delta_\mathrm{D}}
	\end{equation}
	where $I_\mathrm{D}\left(\mathbf{r}\right)$ is the intensity of the optical field, $\bra{\mathrm{g}} \bm{\mu} \ket{\mathrm{d}}$ and $\Delta_\mathrm{D}$ are the transition dipole moment and optical detuning from states $\ket{\mathrm{g}}$ and $\ket{\mathrm{d}}$, and $\epsilon_0$ and $c$ are the vacuum permittivity and speed of light respectively.
	
	A probe laser interrogates the atomic ensemble on the $5\mathrm{S}_{1/2} \rightarrow 5\mathrm{P}_{3/2}$ D2 line, labelled $\ket{\mathrm{g}} \rightarrow \ket{\mathrm{e}}$.
	For an ensemble of atoms in a spatially varying field, such as a focused dipole trap or lattice trap, the spatial dependence of the intensity leads to an apparent broadening of the transition as the probe transition ($\omega_\mathrm{P}$) is perturbed by the light shift of the ground state $\ket{\mathrm{g}}$.
	To correct for this we apply a compensation field ($\omega_\mathrm{C}$) tuned to the $5\mathrm{P}_{3/2} \rightarrow 5\mathrm{D}_{5/2}$ transition, labelled $\ket{\mathrm{e}} \rightarrow \ket{\mathrm{c}}$, that intentionally induces a compensatory light shift in the excited state $\ket{\mathrm{e}}$.
	The total frequency shift of the probe transition frequency is given to first order by the sum of the perturbations due to the dipole and compensation fields:
	\begin{equation}
		\delta\omega_\mathrm{P}\left(\mathbf{r}\right) = \frac{1}{2\epsilon_0 c \hbar^2} \left( \left| \bra{\mathrm{d}} \bm{\mu} \ket{\mathrm{g}} \right|^2 \frac{I_\mathrm{D}\left(\mathbf{r}\right)}{\Delta_\mathrm{D}} - \left| \bra{\mathrm{a}} \bm{\mu} \ket{\mathrm{e}} \right|^2 \frac{I_\mathrm{C}\left(\mathbf{r}\right)}{\Delta_\mathrm{C}}\right)
	\end{equation}
	
	It is evident that, for a given trap field and energy level structure, it is possible to choose an intensity and detuning for the compensation field that eliminates the light shift on probe transition.
	Mode matching the compensation and probe fields automatically compensates for light shifts over the whole beam profile, shown in Figure \ref{fig:EnergyLevels}.b.
	This decouples the motion of the atomic ensemble from line-shape modification of the probe transition, achieving the desired magic regime.
	In order to maintain the compensation it is necessary to stabilise the ratios of $I_\mathrm{D}/\Delta_\mathrm{D}$  and $I_\mathrm{C}/\Delta_\mathrm{C}$.
	In practice this means implementing frequency and power control on both the dipole and compensation lasers.

	One difficulty arises when the desired probe transition involves multiple degenerate Zeeman sub-levels.
	While there may not be a magnetic field to couple to the different angular momentum projection states, the differences in the Clebsch–Gordan coefficients leads to a vector polarizability shift that cannot be simultaneously compensated over all tranisitions between the sub-states. 
	This residual broadening avoided by ensuring that the probe addresses only one specific $\ket{F,m_F} \rightarrow \ket{F',m_F'}$ transition.
	This could be done by applying a strong bias magnetic field to lift the Zeeman state degeneracy, or by pumping atomic population into a single ground state and using an accurately polarised probe beam such that only one excited state is addressed.
	In this experiment we do the latter.
	
\section{Experimental Demonstration}
\begin{figure}
		\centering
		\includegraphics{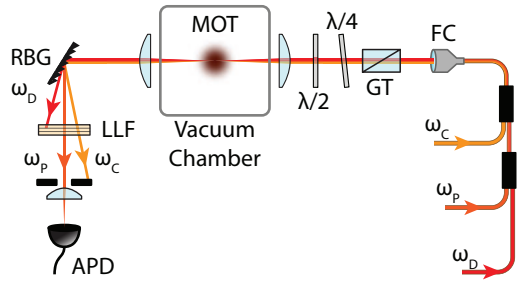}
		\caption{
			\label{fig:Experiment}
			Experimental setup for demonstrating the compensation technique, where RBG is a ruled Bragg grating, LFF is a laser line filter centred at 780 nm, APD is avalanche photo-diode, GT is a Glan-Taylor polarizer, and FC is a fibre collimator.
		}
	\end{figure}

	We demonstrate this technique in a dipole-trapped cold atom ensemble (Figure \ref{fig:Experiment}) that is loaded from a standard three axis magneto optical trap (MOT).
	The dipole field is generated using a titanium-sapphire CW laser that is detuned between \SI{}{\giga\hertz} and \SI{}{\tera\hertz} below the D1 line.
	The light field generates strong trapping forces in the order of several \SI{}{\milli\kelvin}.
	
	The probe and compensation lasers are provided by external cavity diode lasers at \SI{780}{\nano\meter} and \SI{776}{\nano\meter} respectively.
	The probe laser is stabilised to the $^{85}$Rb D2, $\ket{F=3} \rightarrow \ket{F'=4}$ transition and can be tuned up to \SI{72}{\mega\hertz} either side of the transition.
	The compensation laser is coarsely stabilised within \SI{2}{\mega\hertz} to a wavemeter, allowing us to tune over \SI{}{\giga\hertz} via computer control.
	
	We measure the ensemble absorption with a series of 49 light pulses that are \SI{1}{\micro\second} long with \SI{50}{\percent} duty cycle, each pulse stepped in frequency \SI{3}{\mega\hertz} from the preceding pulse, allowing us to scan over \SI{144}{\mega\hertz} in \SI{98}{\micro\second}.
	We can choose to interrogate the ensemble when the dipole laser is on or off.
	In the first case the trapping field is left on during probing while for the latter we switch the trapping field off momentarily.
	
	To achieve good cancellation of the inhomogeneous broadening in the dipole trap there needs to be excellent overlap of all the optical beams.
	This is achieved by combining all the beams within single mode fibre.
	To avoid additional broadening by the partially resolved Zeeman manifolds (as explained earlier) we strongly pump the $\ket{F=3,m_F=+3} \rightarrow \ket{F'=4,m_F'=+4}$ cycling transition on the D2 line and use a very pure right-hand-circular polarisation for the probe.
	The polarisation quality of the dipole and compensation beams is not important as long as there is a non-zero transition dipole moment between the pumped Zeeman sub-states for each field.
	The polarisation tailored combination of fields is then focused through a window into the MOT producing a waist with radius of \SI{27}{\micro\meter}.
	The beam is recollimated after exiting the chamber, and the probe is separated from the other optical fields using a ruled diffraction grating and an etalon filter, and is then collected onto an avalanche photo-diode. 
	
	\begin{figure}[h]
		\centering
		\includegraphics{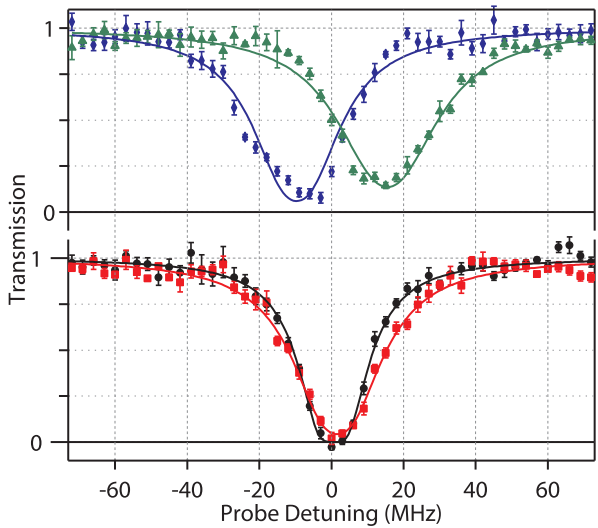}
		\caption{
			\label{fig:Results}
			Transmission spectra of the probe with only the dipole field (green triangles) and only the compensation field (blue diamonds) shown on the top, and with neither field (black circles) and both fields (red squares) on the bottom.
			The compensation field is shown for the optimal detuning calculated.
			Solid lines are used to show fits to the data using the form in eq. \ref{eqn:OD}.
		}
	\end{figure}

	In order to demonstrate the effect of the compensation technique we intentionally apply an extremely strong dipole field using \SI{300}{\milli\watt} of light detuned \SI{250}{\giga\hertz} below the D1 transition, resulting in a frequency shift of $\delta \omega_\mathrm{P} \approx 2\pi\times \SI{14}{\mega\hertz}$.
	This shift and the associated inhomogeneous broadening are clearly visible if one compares the untrapped and unperturbed atoms (black circles, bottom panel) and the trapped atoms (green triangles, top panel) in Figure \ref{fig:Results}.
	To produce a light shift so that  $\delta E^{-}_\mathrm{C} = \delta E^{-}_\mathrm{D}$ we use \SI{1}{\milli\watt} of light in the compensation beam, and calculate a required detuning of $\Delta_\mathrm{C} = \SI{-250}{\mega\hertz}$.
	The compensated spectrum (red squares, bottom panel) in Figure \ref{fig:Results} show that this has successfully returned the lineshape to closely match the unperturbed spectrum.
	Turning off the the trapping field while leaving the compensation on (blue diamonds, top panel) demonstrates that the spectrum has been distorted in the mirror image of that of the trapping field alone, as expected.
	We attribute the small residual broadening present in the compensated spectrum to imperfect population pumping and polarisation selection, resulting in a small fraction of atoms that do not experience the full effect of the compensation.
	
\section{Modelling}
	The interaction between the Rb atoms and the three light fields can be approximated to the four-level system shown in Figure \ref{fig:EnergyLevels}.a.
	This approximation can be made since the guide is sufficiently detuned from state $\ket{\mathrm{d}}$ that the hyperfine splitting is negligible, thus the hyperfine levels can be `averaged' into a single level.
	The same argument holds for the compensation laser and state $\ket{\mathrm{e}}$, while the probe beam only interacts with the $\ket{F=3,m_F=+3} \rightarrow \ket{F'=4,m_F'=+4}$ level.
	The second ground state of Rb is not modelled as neither the dipole, probe or compensation lasers are tuned to this level.
	Furthermore, the probe laser is tuned to the cycling transition of the D2 line, thus negligible decay to the uncoupled ground state is experienced.
	The trapping laser repopulates atoms into both ground states via scattering through D1 transition, however, this effect is slow and any population that does decay to dark ground state during the loading process is pumped with a re-pump laser back to $\ket{\mathrm{g}}$ prior to interrogation. 
	The steady-state absorption of this four-level atomic system is modelled using the density matrix, $\rho$, which is governed by the Liouville equations \cite{RandBook,AuzinshBook} of the general form:
	\begin{equation}
		\label{eqn:Hamiltonian}
		\frac{d\rho}{dt} = -\frac{i}{\hbar}\left[\mathcal{H}_\mathrm{atom}+\mathcal{H}_\mathrm{int},\rho \right]+\Lambda\left(\rho\right)
	\end{equation}
	where $\mathcal{H}_\mathrm{atom}$ and $\mathcal{H}_\mathrm{int}$ are the atomic, and atom-light interaction Hamiltonians under the rotating wave approximation.
	Relaxation is included via the $\Lambda\left(\rho\right)$ term, which accounts for spontaneous emission from the excited states.
	
	As the Stark shift broadening is associated with the motion and position of the atomic sample it is necessary to include a realistic description of these in the model.
	We use a Gaussian beam intensity and fix the waist, $w$, at \SI{27}{\micro\meter} to match the expected spot size at the MOT.
	Within the dipole trap, atoms orbit around the longitudinal axis due to the dipole potential.
	We approximate the spatial variation of the atomic density as a Gaussian peak that is offset from the beam centre by $w/\sqrt{2}$ and has a $1/e$ half-width of $\sim w/3.5$.
	This choice of parameters is based on the modelling of atomic ensembles in traps with depth similar to the kinetic energy of the sample \cite{Hilton2018}. 
	
	The four-level density matrix model was numerically evaluated to produce absorption values for the probe transition for a variety of probe detunings and atomic distances from the centre of the dipole beam.
	The Rabi frequency of the probe, trap, and compensation lasers, along with the detuning of the trap and compensation lasers were held to experimentally measured values and at each radial evaluation point the absorption was scaled with the local atomic density.
	By including the measured physical parameters for our system we are able to test the model against our experimental results.
	
\section{Analysis}
	\begin{figure}
		\centering
		\includegraphics{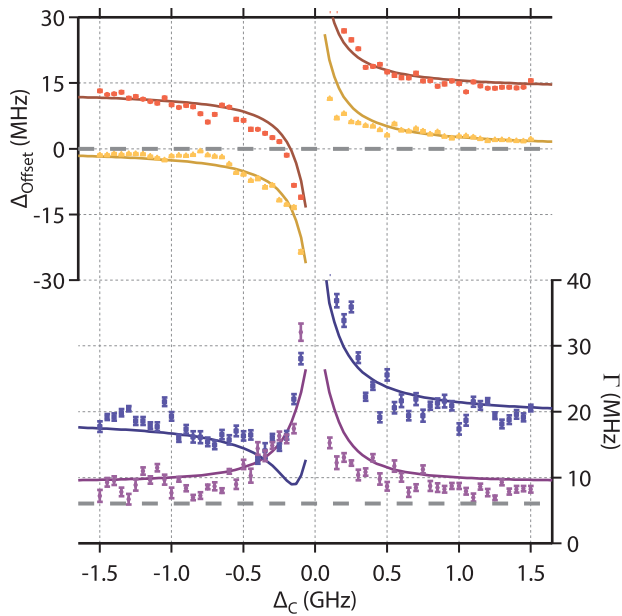}
		\caption{
			\label{fig:Modelling}
			The offset (top) and FWHM (bottom) of the probe transition against the compensation laser detuning for experimental data (markers) and density matrix modelling (solid lines).
			The two data sets are for compensation field only (red circles and purple diamonds), and both compensation and dipole fields (yellow triangles and blue squares), where the grey dashed lines represent the expected offset and FWHM with no additional fields.
		}
	\end{figure}

	We compare the performance of both the experimental and simulated systems by investigating their dependence on the compensation field as the detuning is varied far below, to far above, the direct transition. 
	This allows us to cover the full range of behaviours: (a) under-compensated, (b) optimal detuning for the choice of optical intensities, (c) over-compensation, and (d) anti-compensation when tuned above $\ket{\mathrm{e}} \rightarrow \ket{\mathrm{c}}$.	
		
	We quantify the effects of the trapping and compensation fields on the atomic transition by fitting to the measured optical depth ($\mathrm{OD}$) with the form:
	
	\begin{equation}
		\label{eqn:OD}
		\mathrm{OD}\left(\Delta\right) = \frac{\mathrm{OD}_0}{1+4\left(\left(\Delta-\Delta_\mathrm{offset}\right)/\Gamma\right)^2}
	\end{equation}
	
	where $\mathrm{OD}_0$ is the on-resonance optical depth, $\Delta$ and $\Delta_\mathrm{offset}$ are the frequency detuning and frequency offset from theoretically expected line-centre, and $\Gamma$ is the full width at half maximum (FWHM) of the optical depth.
	We extract the offset and width of the absorption feature over the wide range of $\Delta_\mathrm{c}$ for both the experimental (markers) and modelled system (solid lines), with results shown in Figure \ref{fig:Modelling}.
	We present two data sets for each of the fit parameters: with (blue and red) and without (purple and  yellow) the dipole field turned on during the measurement pulses.
	This allows us to isolate the detrimental effects of the compensation laser from the dipole compensation.
	
	For the frequency offset in the unperturbed case (yellow triangles) we see an avoided crossing feature that is associated with light shift on $\ket{\mathrm{e}}$ by the compensation field: just as one expects from a strong optical field.
	When the dipole field is present (red circles) there is a large constant shift that corresponds to the average potential depth experienced by the atoms.
	As the compensation field is tuned below resonance the frequency offset crosses through zero, demonstrating that the linecentre has been returned to the unperturbed value.
	
	For both these cases the corresponding theory curve shows excellent agreement, both qualitatively and quantitatively.
	This provides confidence in the veracity of the experimental parameters used in the model.

	The width parameter shows more complex behaviour.
	With just the compensation field (purple diamonds) we see broadening as we approach the two-colour resonant excitation due to saturation of the excited states, which approaches the natural line width at large detunings.
	The presence of the dipole field (blue squares) shows a dramatic increase in the spectral width that is highly asymmetric.
	For positive detunings the sign of the ac-Stark shift induced by the compensation field is the same as that for the dipole field resulting in increased broadening, while for negative detunings the two optical fields have opposing two ac-Stark effects on the probe transition, resulting in narrowing.
	This narrowing improves as the compensation field approaches the detuning necessary to completely compensate for the dipole field until it hits a floor set by the width in the compensation field only case.
	We suggest that this is due to a small fraction of the population that remains distributed over a variety of ground $m_F$ states, and is not correctly compensated due to the differential light shift over Zeeman substates.
	
	The theory curves agree well with the experimental data for this parameter although due to the simplified four-level system used the limit set by residual population is not present.
	Instead the system is able to truly approach the natural linewidth at the optimal choice of compensation detuning.

\section{Conclusion}
	We have presented an experimental implementation of the bichromatic magic wavelength technique in a cold rubidium ensemble.
	This is the first demonstration of a concept that has been long theoretically described in the literature.

	We show narrowing of a strongly distorted optical transition, and a compare our results over a large range of compensation laser detunings to a numerically evaluated density matrix model, showing excellent agreement.
	This provides confidence in the implementation of two colour optical dipole traps for long lifetime spectroscopy of any choice of atomic sample.
	
	The technique uses a single additional weak optical field, which allows significant relaxation of the requirements on the energy level structure of the atomic sample.
	This technique opens up a much wider range of possibilities to make use of the revolutionary magic trap techniques, for system where either there is no practical magic wavelength available for the desired probe transition, or where there are limitations on the optical wavelengths that are usable.

\begin{acknowledgments}
	We would like to thank the South Australian government for supporting this research through the PRIF program.
	
	This work was performed in part at the OptoFab node of the Australian National Fabrication Facility utilizing Commonwealth and SA State Government funding.
	
	This research was funded by the Australian government through the Australian Research Council (Project: DE12012028).
	
\end{acknowledgments}

%

\end{document}